\documentclass[twocolumn,amsmath,nofootinbib]{revtex4} 
\begin{document}

\title{Shut up and calculate\footnote{This is the ``director's cut'' version of the September 15 2007 {\it New Scientist} cover story. 
The ``full strength'' version is the much longer article \cite{toe2}, which includes references.}
}

\author{Max Tegmark}
\address{Dept.~of Physics, Massachusetts Institute of Technology, Cambridge, MA 02139}
\begin{abstract}
I advocate an extreme ``shut-up-and-calculate'' approach to physics, where 
our external physical reality is assumed to be purely mathematical.
This brief essay motivates this ``it's all just equations'' assumption and discusses its implications.
\end{abstract}

\maketitle


What is the meaning of life, the universe and everything? In the sci-fi spoof {\it The Hitchhiker's Guide to the 
Galaxy}, the answer was found to be 42; the hardest part turned out to be finding the real question. Indeed, 
although our inquisitive ancestors undoubtedly asked such big questions, their search for a ``theory of 
everything" evolved as their knowledge grew. As the ancient Greeks replaced myth-based explanations with 
mechanistic models of the solar system, their emphasis shifted from asking ``why" to asking ``how".

Since then, the scope of our questioning has dwindled in some areas and mushroomed in others. 
Some questions were abandoned as naive or misguided, such as explaining the sizes of planetary orbits from 
first principles, which was popular during the Renaissance. The same may happen to currently trendy 
pursuits like predicting the amount of dark energy in the cosmos, if it turns out that the amount in our 
neighbourhood is a historical accident. Yet our ability to answer other questions has surpassed earlier 
generations' wildest expectations: Newton would have been amazed to know that we would one day 
measure the age of our universe to an accuracy of 1 per cent, and comprehend the microworld well enough 
to make an iPhone.

Mathematics has played a striking role in these successes. The idea that our universe is in some sense 
mathematical goes back at least to the Pythagoreans of ancient Greece, and has spawned centuries of 
discussion among physicists and philosophers. In the 17th century, Galileo famously stated that the universe 
is a ``grand book" written in the language of mathematics. More recently, the physics Nobel laureate Eugene 
Wigner argued in the 1960s that ``the unreasonable effectiveness of mathematics in the natural sciences" 
demanded an explanation.

Here, I will push this idea to its extreme and argue that our universe is not just described by 
mathematics --- it is mathematics. While this hypothesis might sound rather abstract and far-fetched, it makes 
startling predictions about the structure of the universe that could be testable by observations. It should also 
be useful in narrowing down what an ultimate theory of everything can look like.

The foundation of my argument is the assumption that there exists an external physical reality 
independent of us humans. This is not too controversial: I would guess that the majority of physicists favour 
this long-standing idea, though it is still debated. Metaphysical solipsists reject it flat out, and supporters of 
the so-called Copenhagen interpretation of quantum mechanics may reject it on the grounds that there is no 
reality without observation (New Scientist, 23 June, p 30). Assuming an external reality exists, physics 
theories aim to describe how it works. Our most successful theories, such as general relativity and quantum 
mechanics, describe only parts of this reality: gravity, for instance, or the behaviour of subatomic particles. 
In contrast, the holy grail of theoretical physics is a theory of everything --- a complete description of reality.

My personal quest for this theory begins with an extreme argument about what it is allowed to look 
like. If we assume that reality exists independently of humans, then for a description to be complete, it must 
also be well-defined according to non-human entities --- aliens or supercomputers, say --- that lack any 
understanding of human concepts. Put differently, such a description must be expressible in a form that is 
devoid of any human baggage like ``particle", ``observation" or other English words.

In contrast, all physics theories that I have been taught have two components: mathematical 
equations, and words that explain how the equations are connected to what we observe and intuitively 
understand. When we derive the consequences of a theory, we introduce new concepts --- protons, molecules, 
stars --- because they are convenient. It is important to remember, however, that it is we humans who create 
these concepts; in principle, everything could be calculated without this baggage. For example, a sufficiently 
powerful supercomputer could calculate how the state of the universe evolves over time without interpreting 
what is happening in human terms.

All of this raises the question: is it possible to find a description of external reality that involves no 
baggage? If so, such a description of objects in this external reality and the relations between them would 
have to be completely abstract, forcing any words or symbols to be mere labels with no preconceived 
meanings whatsoever. Instead, the only properties of these entities would be those embodied by the relations 
between them.

This is where mathematics comes in. To a modern logician, a mathematical structure is precisely 
this: a set of abstract entities with relations between them. Take the integers, for instance, or geometric 
objects like the dodecahedron, a favourite of the Pythagoreans. This is in stark 
contrast to the way most of us first perceive mathematics --- either as a sadistic form of punishment, or as a 
bag of tricks for manipulating numbers. Like physics, mathematics has evolved to ask broader questions.

Modern mathematics is the formal study of structures that can be defined in a purely abstract way. 
Think of mathematical symbols as mere labels without intrinsic meaning. It doesn't matter whether you 
write ``two plus two equals four", ``2 + 2 = 4" or ``dos mas dos igual a cuatro". The notation used to denote 
the entities and the relations is irrelevant; the only properties of integers are those embodied by the relations 
between them. That is, we don't invent mathematical structures --- we discover them, and invent only the 
notation for describing them.

So here is the crux of my argument. If you believe in an external reality independent of humans, then 
you must also believe in what I call the mathematical universe hypothesis: that our physical reality is a 
mathematical structure. In other words, we all live in a gigantic mathematical object --- one that is more 
elaborate than a dodecahedron, and probably also more complex than objects with intimidating names like 
Calabi-Yau manifolds, tensor bundles and Hilbert spaces, which appear in today's most advanced theories. 
Everything in our world is purely mathematical --- including you.

If that is true, then the theory of everything must be purely abstract and mathematical. Although we 
do not yet know what the theory would look like, particle physics and cosmology have reached a point 
where all measurements ever made can be explained, at least in principle, with equations that fit on a few 
pages and involve merely 32 unexplained numerical constants (Physical Review D, vol 73, 023505). So it 
seems possible that the correct theory of everything could even turn out to be simple enough to describe with 
equations that fit on a T-shirt.

Before discussing whether the mathematical universe hypothesis is correct, however, there is a more 
urgent question: what does it actually mean? To understand this, it helps to distinguish between two ways of 
viewing our external physical reality. One is the outside overview of a physicist studying its mathematical 
structure, like a bird surveying a landscape from high above; the other is the inside view of an observer 
living in the world described by the structure, like a frog living in the landscape surveyed by the bird.

One issue in relating these two perspectives involves time. A mathematical structure is by definition 
an abstract, immutable entity existing outside of space and time. If the history of our universe were a movie, 
the structure would correspond not to a single frame but to the entire DVD. So from the bird's perspective, 
trajectories of objects moving in four-dimensional space-time resemble a tangle of spaghetti. Where the frog 
sees something moving with constant velocity, the bird sees a straight strand of uncooked spaghetti. Where 
the frog sees the moon orbit the Earth, the bird sees two intertwined spaghetti strands. To the frog, the world 
is described by Newton's laws of motion and gravitation. To the bird, the world is the geometry of the pasta.

A further subtlety in relating the two perspectives involves explaining how an observer could be 
purely mathematical. In this example, the frog itself must consist of a thick bundle of pasta whose highly 
complex structure corresponds to particles that store and process information in a way that gives rise to the 
familiar sensation of self-awareness.

Fine, so how do we test the mathematical universe hypothesis? For a start, it predicts that further 
mathematical regularities remain to be discovered in nature. Ever since Galileo promulgated the idea of a 
mathematical cosmos, there has been a steady progression of discoveries in that vein, including the standard 
model of particle physics, which captures striking mathematical order in the microcosm of elementary 
particles and the macrocosm of the early universe.

That's not all, however. The hypothesis also makes a more dramatic prediction: the existence of 
parallel universes. Many types of ``multiverse" have been proposed over the years, and it is useful to classify 
them into a four-level hierarchy. The first three levels correspond to non-communicating parallel worlds 
within the same mathematical structure: level I simply means distant regions from which light has not yet 
had time to reach us; level II covers regions that are forever unreachable because of the cosmological 
inflation of intervening space; and level III, often called ``many worlds", involves non-communicating parts 
of the so-called Hilbert space of quantum mechanics into which the universe can ``split" during certain 
quantum events. Level IV refers to parallel worlds in distinct mathematical structures, which may have 
fundamentally different laws of physics.

Today's best estimates suggest that we need a huge amount of information, perhaps a Googol ($10^{100}$) bits, to 
fully describe our frog's view of the observable universe, down to the positions of every star and grain of 
sand. Most physicists hope for a theory of everything that is much simpler than this and can be specified in 
few enough bits to fit in a book, if not on a T-shirt. The mathematical universe hypothesis implies that such 
a simple theory must predict a multiverse. Why? Because this theory is by definition a complete description 
of reality: if it lacks enough bits to completely specify our universe, then it must instead describe all possible 
combinations of stars, sand grains and such --- so that the extra bits that describe our universe simply encode 
which universe we are in, like a multiversal telephone number. In this way, describing a multiverse can be 
simpler than describing a single universe.

Pushed to its extreme, the mathematical universe hypothesis implies the level-IV multiverse, which 
includes all the other levels within it. If there is a particular mathematical structure that is our universe, and 
its properties correspond to our physical laws, then each mathematical structure with different properties is 
its own universe with different laws. Indeed, the level-IV multiverse is compulsory, since mathematical 
structures are not ``created" and don't exist ``somewhere" --- they just exist. Stephen Hawking once asked, 
``What is it that breathes fire into the equations and makes a universe for them to describe?" In the case of 
the mathematical cosmos, there is no fire-breathing required, since the point is not that a mathematical 
structure describes a universe, but that it is a universe. 

The existence of the level-IV multiverse also answers a confounding question emphasised by the 
physicist John Wheeler: even if we found equations that describe our universe perfectly, then why these 
particular equations, not others? The answer is that the other equations govern parallel universes, and that 
our universe has these particular equations because they are statistically likely, given the distribution of 
mathematical structures that can support observers like us.

It is crucial to ask whether parallel universes are within the purview of science, or are merely 
speculation. Parallel universes are not a theory in themselves, but rather a prediction made by certain 
theories. For a theory to be falsifiable, we need not be able to observe and test all its predictions, merely at 
least one of them. General relativity, for instance, has successfully predicted many things that we can 
observe, such as gravitational lensing, so we also take seriously its predictions for things we cannot, like the 
internal structure of black holes.

So here's a testable prediction of the mathematical universe hypothesis: if we exist in many parallel 
universes, then we should expect to find ourselves in a typical one.  Suppose we succeed in computing the 
probability distribution for some quantity, say the dark energy density or the dimensionality of space, 
measured by a typical observer in the part of the multiverse where this quantity is defined. If we find that 
this distribution makes the value measured in our own universe highly atypical, it would rule out the 
multiverse, and hence the mathematical universe hypothesis.
Although we are still far from understanding the requirements for life, we could start testing the multiverse 
prediction by assessing how typical our universe is as regards dark matter, dark energy and neutrinos, 
because these substances affect only better understood processes like galaxy formation. This prediction has 
passed the first of such tests, because the abundance of these substances has been measured to be rather 
typical of what you might measure from a random galaxy in a multiverse. However, more accurate 
calculations and measurements might still rule out such a multiverse.

Ultimately, why should we believe the mathematical universe hypothesis? Perhaps the most 
compelling objection is that it feels counter-intuitive and disturbing. I personally dismiss this as a failure to 
appreciate Darwinian evolution. Evolution endowed us with intuition only for those aspects of physics that 
had survival value for our distant ancestors, such as the parabolic trajectories of flying rocks. Darwin's 
theory thus makes the testable prediction that whenever we look beyond the human scale, our evolved 
intuition should break down.

We have repeatedly tested this prediction, and the results overwhelmingly support it: our intuition 
breaks down at high speeds, where time slows down; on small scales, where particles can be in two places at 
once; and at high temperatures, where colliding particles change identity. To me, an electron colliding with a 
positron and turning into a Z-boson feels about as intuitive as two colliding cars turning into a cruise ship. 
The point is that if we dismiss seemingly weird theories out of hand, we risk dismissing the correct theory of 
everything, whatever it may turn out to be.

If the mathematical universe hypothesis is true, then it is great news for science, allowing the 
possibility that an elegant unification of physics and mathematics will one day allow us to understand reality 
more deeply than most dreamed possible. Indeed, I think the mathematical cosmos with its multiverse is the 
best theory of everything that we could hope for, because it would mean that no aspect of reality is off-limits 
from our scientific quest to uncover regularities and make quantitative predictions.

However, it would also shift the ultimate question about the universe once again. We would abandon 
as misguided the question of which particular mathematical equations describe all of reality, and instead ask 
how to compute the frog's view of the universe --- our observations --- from the bird's view. That would 
determine whether we have uncovered the true structure of our universe, and help us figure out which corner 
of the mathematical cosmos is our home.

\end{document}